\documentclass[11pt]{article}
\usepackage[utf8]{inputenc}
\usepackage[T1]{fontenc}
\usepackage{lmodern}
\usepackage{textcomp}
\usepackage[margin=1in]{geometry}
\usepackage{microtype}
\usepackage{amsmath,amssymb}
\usepackage{enumitem}
\usepackage{graphicx}
\usepackage{float}
\usepackage{booktabs,tabularx,array}
\newcolumntype{L}[1]{>{\raggedright\arraybackslash}p{#1}}

\usepackage{tikz}
\usetikzlibrary{positioning}
\usepackage[hidelinks]{hyperref}

\title{\textbf{Zero-Knowledge Extensions on Solana: \\ A Theory of ZK Architecture}}
\author{Jotaro Yano \\ \small Independent Researcher, Japan \\ \href{mailto:jotaro.yano@jotaro-yano.org}{\texttt{jotaro.yano@jotaro-yano.org}}}
\date{October 23, 2025}

\begin{document}
\maketitle

\begin{abstract}
\noindent This paper reconstructs zero-knowledge extensions on Solana as an architecture theory. Drawing on the existing ecosystem and on the author’s prior papers and implementations as reference material, we propose a two-axis model that normalizes zero-knowledge (ZK) use by purpose (scalability vs.\ privacy) and by placement (on-chain vs.\ off-chain). On this grid we define five layer-crossing invariants---origin authenticity, replay-safety, finality alignment, parameter binding, and private consumption---which serve as a common vocabulary for reasoning about correctness across modules and chains. The framework covers the Solana Foundation’s three pillars (ZK Compression, Confidential Transfer, light clients/bridges) together with surrounding components (Light Protocol/Helius, Succinct SP1, RISC Zero, Wormhole, Tinydancer, Arcium). From the theory we derive two design abstractions---Proof-Carrying Message (PCM) and a Verifier Router Interface---and a cross-chain counterpart, Proof-Carrying Interchain Message (PCIM), indicating concrete avenues for extending the three pillars.
\par\vspace{1em}
\end{abstract}

\newpage

\section{Introduction}
The animating principle of public blockchains is that users can verify for themselves. At scale, however, this principle meets two persistent forms of friction: the computational and state burden of full verification, and the privacy leakage induced by ubiquitous transparency. Zero-knowledge (ZK) techniques have emerged as the canonical mediator between these forces. They compress verification through succinct proofs and permit selective non-disclosure without abandoning public verifiability.

Solana’s trajectory emphasizes keeping scalability and composability at L1, enriching the base layer with primitives rather than outsourcing correctness wholesale to external domains. ZK Compression illustrates this stance: compression is treated as a first-class L1 construct that reduces state surface while maintaining atomic composability \hyperlink{r13}{[13]}, \hyperlink{r14}{[14]}, \hyperlink{r16}{[16]}. At the same time, the broader ZK surface on Solana remains a moving target. Confidential Transfer has undergone redesign; official documentation and workflows are available and continue to evolve \hyperlink{r7}{[7]}; Light-clients and bridge efforts are evolving \hyperlink{r15}{[15]}, \hyperlink{r5}{[5]}; and multiple stacks---zkVM receipts, privacy L2s, message-attestation transports---coexist with heterogeneous interfaces and security assumptions. A unified architectural vocabulary is needed to compare these efforts, identify gaps, and guide extensions without overfitting to any one implementation.

This paper views ZK not as a monolith but as a cryptographic interface among layers and modules. We contribute a two-axis classification of ZK use by purpose (scalability vs.\ privacy) and placement (on-chain vs.\ off-chain), yielding four canonical quadrants that cover current practice---compressed accounts, confidential transfers, zkVM receipts, and private L2s/MPC networks. On this grid we introduce five invariants---origin authenticity, replay-safety, finality alignment, parameter binding, private consumption---which we claim form the right unit of discourse for ZK-enabled systems that cross consensus boundaries.

\paragraph{Contributions.}
\begin{enumerate}[leftmargin=1.15em]
  \item A ZK architecture theory for Solana based on the two-axis model and five invariants, intended to normalize discourse across stacks.
  \item An analysis of representative patterns---cross-domain private execution and on-chain general verification---using the framework, with the author’s prior results used only as references alongside ecosystem implementations \hyperlink{r1}{[1]}, \hyperlink{r2}{[2]}, \hyperlink{r4}{[4]}.
  \item Design propositions that extend the Foundation’s three pillars while remaining implementation-agnostic: Proof-Carrying Messages (PCM) for composable compressed-state updates; Proof-Carrying Interchain Messages (PCIM) for bridge-safe message semantics; and a Verifier Router Interface that decouples applications from proof systems.
\end{enumerate}

\section{Background: ZK on Solana Today (Roadmap and Ecosystem)}
\paragraph{Roadmap.} The Solana Foundation’s ZK strategy can be read as three pillars. (A) Scalability centers ZK Compression as an L1 primitive: application state is represented in compressed form with succinct validity checks, preserving L1 atomic composability \hyperlink{r13}{[13]}, \hyperlink{r14}{[14]}. By shifting large state off hot storage while verifying small proofs on-chain, the approach targets sustainable scale without splitting execution across rollups. (B) Privacy and compliance focuses on confidential functionality---most notably Confidential Transfer---with a renewed emphasis on selective disclosure and auditability following recent redesigns; official documentation and workflows are available and continue to evolve \hyperlink{r7}{[7]}. (C) Verifiability and interoperability pushes light clients and bridges, aiming to extend ``verify for yourself'' to cross-chain settings and minimize reliance on off-chain oracles \hyperlink{r15}{[15]}, \hyperlink{r5}{[5]}.

\begin{figure}[H]
\centering
\begin{tikzpicture}[node distance=14mm, every node/.style={draw, rounded corners, inner sep=6pt, align=center}]
  \node (A) {\shortstack[c]{Scalability\\{\footnotesize ZK Compression}}};
  \node (B) [right=of A] {\shortstack[c]{Privacy \& Compliance\\{\footnotesize Confidential Transfer}}};
  \node (C) [right=of B] {\shortstack[c]{Verifiability \& Interoperability\\{\footnotesize Light clients / bridges}}};
\end{tikzpicture}
\caption{Solana Foundation ZK roadmap: three pillars (simple schematic).}
\end{figure}
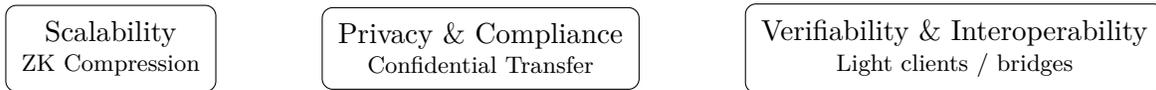

\paragraph{Ecosystem.} These pillars are instantiated and complemented by several projects. \textit{Light Protocol} implements the core mechanics of ZK Compression \hyperlink{r13}{[13]}; \textit{Helius} supplies indexing and distribution tooling that make compressed state operationally observable \hyperlink{r14}{[14]}. \textit{Succinct SP1} and \textit{RISC Zero} package arbitrary off-chain execution into small receipts with on-chain verifiers and router abstractions, enabling applications to check general computation at L1 \hyperlink{r10}{[10]}, \hyperlink{r9}{[9]}. For cross-domain delivery, \textit{Wormhole} provides verifiable message approvals (VAAs) that capture origin authenticity under a threshold-signature model \hyperlink{r5}{[5]}; \textit{Aztec} exposes inbox/portal interfaces through which messages can be parameter-bound and privately consumed via commitments and nullifiers \hyperlink{r6}{[6]}. \textit{Tinydancer} explores Solana light clients \hyperlink{r15}{[15]}. \textit{Arcium} targets encrypted and private computation using MPC with optional ZK attestations \hyperlink{r12}{[12]}. We treat these systems as exemplars; the theory abstracts their guarantees and failure modes into a common vocabulary, enabling principled comparison and composition.

\section{A Theory of ZK Architecture: Use-Model and Five Invariants}
\subsection{Two-Axis Model}
We classify ZK use by purpose and placement. The following table summarizes the four quadrants and examples.

\begin{table}[H]
\centering
\caption{Two-axis use-model: purpose and placement with examples.}
\renewcommand{\arraystretch}{1.15}
\begin{tabularx}{\linewidth}{L{0.22\linewidth} L{0.26\linewidth} L{0.46\linewidth}}
\toprule
\textbf{Quadrant} & \textbf{Placement / Purpose} & \textbf{Typical examples} \\
\midrule
On-chain $\times$ scalability & On-chain / scalability & Small proofs verified on L1 (e.g., SNARK/zkVM receipts \hyperlink{r18}{[18]}, \hyperlink{r17}{[17]}). \\
On-chain $\times$ privacy     & On-chain / privacy     & Confidential transfers (privacy proofs verified on L1 \hyperlink{r7}{[7]}). \\
Off-chain $\times$ scalability& Off-chain / scalability& ZK coprocessors executing heavy logic and returning succinct proofs (SP1, RISC Zero \hyperlink{r10}{[10]}, \hyperlink{r9}{[9]}). \\
Off-chain $\times$ privacy    & Off-chain / privacy    & Private L2s / MPC networks; commitments, nullifiers, selective receipts (Aztec, Arcium \hyperlink{r6}{[6]}, \hyperlink{r12}{[12]}). \\
\bottomrule
\end{tabularx}
\end{table}

\paragraph{Narrative explanation.}
The two-axis model is descriptive rather than prescriptive. Many real deployments straddle quadrants: zkVM-based coprocessors (off-chain execution) whose receipts are checked on L1 (on-chain verification) \hyperlink{r10}{[10]}, \hyperlink{r9}{[9]}, or confidential assets that interoperate with compressed accounts to achieve both rent reduction and privacy \hyperlink{r13}{[13]}, \hyperlink{r14}{[14]}, \hyperlink{r7}{[7]}. The model’s value is diagnostic: it makes explicit which guarantees are held on-chain and which are deferred to off-chain components. In the Solana context, Light Protocol/Helius instantiate the \emph{on-chain $\times$ scalability} quadrant \hyperlink{r13}{[13]}, \hyperlink{r14}{[14]}; Confidential Transfer inhabits \emph{on-chain $\times$ privacy} \hyperlink{r7}{[7]}; zkVM stacks such as Succinct SP1 and RISC Zero enable hybrids where proving is off-chain but verification is on-chain \hyperlink{r10}{[10]}, \hyperlink{r9}{[9]}; Aztec and Arcium populate \emph{off-chain $\times$ privacy} with private consumption and MPC-backed computation \hyperlink{r6}{[6]}, \hyperlink{r12}{[12]}; Wormhole provides transport that connects these quadrants by carrying authenticated messages \hyperlink{r5}{[5]}; and Tinydancer explores how minimal on-chain verifiers can reason about remote state \hyperlink{r15}{[15]}.

\subsection{Five Invariants}
The following table records the five invariants and typical enforcing layers.

\begin{table}[H]
\centering
\caption{Five invariants and typical enforcement layers.}
\renewcommand{\arraystretch}{1.15}
\begin{tabularx}{\linewidth}{L{0.22\linewidth} L{0.54\linewidth} L{0.20\linewidth}}
\toprule
\textbf{Invariant} & \textbf{Brief definition} & \textbf{Typical enforcing layer(s)} \\
\midrule
Origin authenticity & Receiver can verify the sender identity under a stated signing assumption. & Transport attestation / portal / L1 \\
Replay-safety & Single-use acceptance per message identifier; duplicates or reorders fail. & Receiver-side portal / L1 \\
Finality alignment & Acceptance respects the sender’s consensus finality predicate. & Receiver policy / bridge / L1 \\
Parameter binding & Off-chain parameters are bound to the message; substitution/frontrunning is prevented. & Commitment in message; L2 inbox / portal / L1 \\
Private consumption & Consumption gated by secret knowledge (or a witness), with controlled disclosure. & Privacy L2 / MPC layer \\
\bottomrule
\end{tabularx}
\end{table}

\paragraph{Explanatory notes.}
The invariants are allocation targets: a design specifies where each property is guaranteed. For example, with Wormhole-style VAAs and an Aztec-like inbox, origin authenticity is anchored by transport attestation, replay-safety and finality alignment are enforced by a receiver portal that locks identifiers and honors source finality, while parameter binding and private consumption are realized by commitments and secret openings at the privacy layer \hyperlink{r5}{[5]}, \hyperlink{r6}{[6]}. In zkVM receipt flows (SP1 or RISC Zero), parameter binding is conveyed via the public-input encoding and verified on-chain, while application-level identifiers implement replay-safety \hyperlink{r10}{[10]}, \hyperlink{r9}{[9]}. For ZK Compression (Light Protocol with Helius observability), updates can be rephrased as proof-carrying messages that combine origin, replay, and binding with succinct validity, leaving finality to L1 acceptance rules \hyperlink{r13}{[13]}, \hyperlink{r14}{[14]}.

\section{Reference Pattern I: Cross-Domain Private Execution}
A canonical ``ZK coprocessor'' pattern proceeds in stages. A Solana program emits a request; a transport (for example, a VAA-like attestation) supplies origin authenticity and carries structure sufficient for replay-safety. The receiving portal enforces finality alignment and injects a parameter-bound message---typically a commitment to a secret concatenated with public parameters---into a privacy-preserving environment such as a private L2 inbox. The destination privately consumes the message by opening the secret, producing a result whose correctness can, when desired, be summarized by a succinct receipt for later on-chain verification \hyperlink{r5}{[5]}, \hyperlink{r6}{[6]}, \hyperlink{r1}{[1]}, \hyperlink{r2}{[2]}.

This is an off-chain $\times$ privacy construction that nevertheless anchors semantics at L1: origin, replay, and finality are enforced at acceptance boundaries; parameter binding and consumption are enforced in the privacy layer; and proof verification can return to L1. Existing stacks instantiate variants of this pattern; the articulation here isolates the invariant allocation so that implementers can reason about safety regardless of transport or L2 choice.

\section{Reference Pattern II: On-Chain General Verification}
A second pattern is general verification on L1. Two routes predominate:
\begin{itemize}[leftmargin=1.15em]
  \item \textbf{Receipt route.} Off-chain execution (often in a zkVM) is packaged as a small proof with a stable public-input interface; L1 verifies the receipt \hyperlink{r10}{[10]}, \hyperlink{r9}{[9]}, \hyperlink{r18}{[18]}, \hyperlink{r17}{[17]}.
  \item \textbf{Transparent/PQ route.} Proof systems with transparent setup (for example, STARKs) and, optionally, post-quantum signatures are used to favor long-horizon auditability and setup independence, with heavier verification costs \hyperlink{r19}{[19]}, \hyperlink{r20}{[20]}, \hyperlink{r21}{[21]}, \hyperlink{r4}{[4]}, \hyperlink{r3}{[3]}.
\end{itemize}
The framework recommends a Verifier Router Interface that presents a single application-level interface, for example, \texttt{verify(proof, public\_values, vk\_id)}, while allowing operators to select the underlying proof system per deployment. Invariants are allocated explicitly: origin/finality via L1 acceptance rules; parameter binding via the encoding of \texttt{public\_values}; replay-safety via application identifiers; private consumption layered as needed. The router makes proof-system substitution, aggregation, and evolution manageable without rewriting application logic.

\section{Extending the Three Pillars: Design Propositions}
\subsection{Scalability: Proof-Carrying Messages (PCM) for Composable Compression}
We propose Proof-Carrying Messages for compressed-state updates. A PCM couples an update command with a validity proof and the identifiers necessary for origin and replay checks. The receiver verifies: (i) origin and single-use semantics; (ii) that the command satisfies the transition relation (parameter binding); and (iii) that pre- and post-roots are consistent (finality alignment). PCMs support batching and third-party distribution (for example, verifiable airdrops), elevating compression from a storage format to a verifiable update protocol that fits Solana’s composability \hyperlink{r13}{[13]}, \hyperlink{r14}{[14]}.

\subsection{Privacy and Compliance: Confidential Asset Interface and ZK-of-MPC}
For confidential functionality, we advocate an asset-level interface that specifies reversible transparency \,$\leftrightarrow$\, privacy and role-based selective disclosure. Off-chain private computation---especially MPC networks---can be integrated via ZK-of-MPC: perform the computation privately, then prove only result correctness to L1. Invariants partition naturally: private consumption off-chain; origin/replay/finality/parameter binding on L1 and at bridge boundaries \hyperlink{r12}{[12]}, \hyperlink{r5}{[5]}, \hyperlink{r6}{[6]}. The result is a design that preserves audit trails while meeting confidentiality constraints.

\subsection{Verifiability and Interoperability: PCIM and a Common Verification Sink}
We introduce Proof-Carrying Interchain Messages. A PCIM embeds a finality tag, a single-use identifier, and a parameter commitment so that any receiver can check origin, enforce replay-safety, and verify parameter binding mechanically, independent of bridge internals. In parallel, the Verifier Router Interface provides a common sink at L1 for receipts from heterogeneous proving systems. PCIM plus the router transform ZK bridges and light clients into commuting diagrams over the invariants, clarifying how correctness composes across domains \hyperlink{r5}{[5]}, \hyperlink{r10}{[10]}, \hyperlink{r9}{[9]}.

\section{Model and Security (Sketch)}
Let $\mathbb{M}$ be a message space, $\mathbb{I}$ an identifier space, and $\mathcal{C}$ a commitment space. Let $\mathrm{Com} : \{0,1\}^\ast \to \mathcal{C}$ be a binding commitment and $\mathrm{Acc} : \mathbb{M} \times \mathcal{C} \times \mathbb{I} \to \{0,1\}$ an acceptance predicate.
\begin{itemize}[leftmargin=1.15em]
  \item \textbf{Origin authenticity} requires completeness/soundness of a threshold (or multi-sig) authentication scheme for messages in $\mathbb{M}$ \hyperlink{r5}{[5]}.
  \item \textbf{Replay-safety} requires that for any PPT adversary, $\mathrm{Acc}(m,c,i)=1$ occurs at most once per $i \in \mathbb{I}$, except with negligible probability.
  \item \textbf{Finality alignment} requires that $\mathrm{Acc}$ accepts only messages attested under a sender-side finality predicate $F$; acceptance from non-final observations is negligible \hyperlink{r22}{[22]}.
  \item \textbf{Parameter binding} requires that if $\mathrm{Acc}(m,\mathrm{Com}(\mathrm{params}), i)=1$, then $(m,\mathrm{params})$ satisfies a declared relation $R$; an adversary cannot cause acceptance for $(m',\mathrm{params}')$ linked to the same $i$.
  \item \textbf{Private consumption} requires that acceptance implies knowledge (or extractability) of a witness for $R$ by the consumer; transcripts reveal no function of the secret beyond what $R$ permits \hyperlink{r6}{[6]}, \hyperlink{r12}{[12]}.
\end{itemize}
A PCM is a tuple $(m, \mathrm{Com}(\mathrm{params}), i)$ with a proof of $R$ and identifiers enabling replay checks. A PCIM additionally carries a sender-finality tag and transport-level attestation. Under standard assumptions (EUF-CMA signatures, binding commitments, knowledge-sound proofs), one can state compositionality claims: (i) disjoint PCMs/PCIMs compose without violating replay or binding; (ii) relays that preserve $(m, \mathrm{Com}(\mathrm{params}), i)$ cannot introduce substitution attacks; (iii) batching preserves acceptance if and only if $R$ is closed under the batch operator. Full proofs are out of scope; the goal here is to fix the interfaces and predicates so that such proofs can be developed.

\section{Related Work (Expanded)}
\paragraph{ZK foundations.} Pairing-based SNARKs (e.g., Groth16) provide extremely small proofs and fast verification, at the cost of setup and algebraic assumptions \hyperlink{r18}{[18]}, \hyperlink{r17}{[17]}. Transparent proof systems (e.g., STARKs) avoid trusted setup and track well to post-quantum concerns but impose heavier verification \hyperlink{r19}{[19]}, \hyperlink{r20}{[20]}, \hyperlink{r21}{[21]}. Polynomial-commitment schemes (KZG, IPA), Fiat--Shamir transforms, sum-check/FRI, and incrementally verifiable computation underpin modern recursive and aggregation pipelines.

\paragraph{General-purpose zkVM systems.} \textit{Succinct SP1} and \textit{RISC Zero} package arbitrary programs into proofs/receipts suitable for on-chain verification and expose router-style verifiers that decouple applications from proof formats \hyperlink{r10}{[10]}, \hyperlink{r9}{[9]}. Audited verifier routers and universal verifiers provide the template for a Verifier Router Interface on Solana.

\paragraph{Compression and observability.} \textit{Light Protocol} implements ZK Compression as an L1 primitive \hyperlink{r13}{[13]}, while \textit{Helius} contributes indexing and distribution tooling that make compressed state operationally observable \hyperlink{r14}{[14]}. These systems illustrate how compression can migrate from a storage format to a verifiable update protocol via proof-carrying messages.

\paragraph{Interoperability and clients.} \textit{Wormhole} (VAA-based message attestation) offers a practical basis for origin authenticity in cross-domain flows \hyperlink{r5}{[5]}. \textit{Aztec} provides inbox/portal patterns for parameter binding and private consumption \hyperlink{r6}{[6]}. \textit{Tinydancer} explores Solana light clients and proof-window designs that reduce reliance on off-chain RPC while enabling minimal on-chain verification \hyperlink{r15}{[15]}. Outside Solana, \textit{Axiom} shows how succinct on-chain queries over historical data can be packaged, informing Solana-side coprocessor designs \hyperlink{r11}{[11]}.

\paragraph{Privacy computation.} \textit{Arcium} targets encrypted and private computation using MPC as the primary engine and ZK attestations as an audit layer, aligning with ZK-of-MPC strategies that reconcile confidentiality with on-chain auditability \hyperlink{r12}{[12]}. Solana-native PQZK efforts---full-chain PQC+STARK verification---demonstrate feasibility on L1 \hyperlink{r3}{[3]}, \hyperlink{r4}{[4]}.

\section{Discussion and Limitations}
The framework deliberately abstracts away concrete parameters---proof sizes, public-input carriage, fee markets---and network constraints that matter in deployment. Such details must be supplied per system. Operational assumptions (guardian-set security, indexer availability, key-management hygiene) sit outside the five invariants yet influence realized safety; likewise, bridge governance and fault domains are orthogonal. PCIM can standardize message-level safety, but it does not eliminate institutional risk.

The paper also does not prescribe a single proof system. The Verifier Router mitigates proof-format lock-in, but its correctness hinges on high-quality audits and careful key/version management. Finally, privacy properties depend on application-level data flows; private consumption constrains only the acceptance event, not all side channels.

\section{Conclusion}
This work presented a ZK architecture theory for Solana that organizes heterogeneous practice into a two-axis model and five invariants, enabling designs to be viewed as allocation matrices over layers and domains. Casting cross-domain private execution and on-chain general verification in these terms elevates them from implementation patterns to principled constructions. The proposed PCM/PCIM abstractions and Verifier Router Interface provide concrete avenues to extend the Foundation’s pillars: composable compressed-state updates, confidential assets with selective disclosure and on-chain auditability, and interoperable verification sinks for receipts of many kinds. Future work includes full game-based formalizations and proofs for the invariants, and collaboration with ecosystem stakeholders to refine these abstractions into actionable specifications suited for standardization.


\begin{thebibliography}{99}

\hypertarget{r1}{}\bibitem{zk-coprocessor-bridge}
PQZK Labs.
\newblock \emph{zk-coprocessor-bridge}, 2025.
\newblock Version 0.1.0; accessed 2025-10-21.
\newblock \url{https://github.com/pqzk-labs/zk-coprocessor-bridge}.

\hypertarget{r2}{}\bibitem{yano_2025_17409587}
J.~Yano.
\newblock ZK Coprocessor Bridge: Replay-Safe Private Execution from Solana to Aztec via Wormhole.
\newblock Zenodo, 2025.
\newblock doi: \href{https://doi.org/10.5281/zenodo.17409587}{10.5281/zenodo.17409587}.

\hypertarget{r3}{}\bibitem{solana_pqzk_fullchain}
PQZK Labs.
\newblock \emph{solana-pqzk-fullchain}, 2025.
\newblock Version v0.1.0.
\newblock \url{https://github.com/pqzk-labs/solana-pqzk-fullchain}.

\hypertarget{r4}{}\bibitem{cryptoeprint:2025/1741}
J.~Yano.
\newblock Full L1 On-Chain ZK-STARK+PQC Verification on Solana: A Measurement Study.
\newblock Cryptology ePrint Archive, Paper 2025/1741, 2025.
\newblock \url{https://eprint.iacr.org/2025/1741}.

\hypertarget{r5}{}\bibitem{wormhole_docs}
Wormhole Foundation.
\newblock Wormhole Documentation, 2025.
\newblock accessed 2025-10-21.
\newblock \url{https://docs.wormhole.com/}.

\hypertarget{r6}{}\bibitem{aztec_docs}
Aztec Labs.
\newblock Aztec Protocol Documentation, 2025.
\newblock accessed 2025-10-21.
\newblock \url{https://docs.aztec.network/}.

\hypertarget{r7}{}\bibitem{solana_docs}
Solana Foundation.
\newblock Solana Documentation, 2025.
\newblock accessed 2025-10-21.
\newblock \url{https://solana.com/docs}.

\hypertarget{r8}{}\bibitem{anchor_lang_crate}
Solana Foundation.
\newblock anchor-lang: Solana Program Framework, 2025.
\newblock v0.31.1; accessed 2025-10-21.
\newblock \url{https://crates.io/crates/anchor-lang}.

\hypertarget{r9}{}\bibitem{risc0_docs}
RISC Zero.
\newblock RISC Zero zkVM Documentation, 2025.
\newblock accessed 2025-10-21.
\newblock \url{https://docs.risczero.com/}.

\hypertarget{r10}{}\bibitem{succinct_docs}
Succinct Labs.
\newblock Succinct Documentation, 2025.
\newblock accessed 2025-10-21.
\newblock \url{https://docs.succinct.xyz/}.

\hypertarget{r11}{}\bibitem{axiom_docs}
Axiom Labs.
\newblock Axiom Documentation, 2025.
\newblock accessed 2025-10-21.
\newblock \url{https://docs.axiom.xyz/}.

\hypertarget{r12}{}\bibitem{arcium_docs}
Arcium.
\newblock Arcium Documentation, 2025.
\newblock accessed 2025-10-21.
\newblock \url{https://docs.arcium.com/}.

\hypertarget{r13}{}\bibitem{light_protocol_whitepaper}
Light Protocol Team.
\newblock Scaling the Design Space for On-chain Applications with ZK Compression (Whitepaper), 2025.
\newblock accessed 2025-10-21.
\newblock \url{https://www.zkcompression.com/references/whitepaper}.

\hypertarget{r14}{}\bibitem{helius_zkcompression_api}
Helius.
\newblock ZK Compression API Documentation, 2025.
\newblock accessed 2025-10-23.
\newblock \url{https://www.helius.dev/docs/api-reference/zk-compression}.

\hypertarget{r15}{}\bibitem{tinydancer_site}
Tinydancer.
\newblock Tinydancer: The First Light Client on Solana, 2025.
\newblock accessed 2025-10-23.
\newblock \url{https://www.tinydancer.io/}.

\hypertarget{r16}{}\bibitem{yakovenko_solana_wp}
A.~Yakovenko.
\newblock Solana: A New Architecture for a High Performance Blockchain (Whitepaper), 2018.
\newblock accessed 2025-10-21.
\newblock \url{https://solana.com/solana-whitepaper.pdf}.

\hypertarget{r17}{}\bibitem{cryptoeprint:2019/953}
A.~Gabizon, Z.~J.~Williamson, O.~Ciobotaru.
\newblock PLONK: Permutations over Lagrange-bases for Oecumenical Noninteractive arguments of Knowledge.
\newblock Cryptology ePrint Archive, Paper 2019/953, 2019.
\newblock \url{https://eprint.iacr.org/2019/953}.

\hypertarget{r18}{}\bibitem{groth_2016}
J.~Groth.
\newblock On the Size of Pairing-Based Non-interactive Arguments.
\newblock In \emph{EUROCRYPT 2016}, pp.~305--326, Springer, 2016.
\newblock doi: \href{https://doi.org/10.1007/978-3-662-49896-5_11}{10.1007/978-3-662-49896-5\_11}.

\hypertarget{r19}{}\bibitem{fri_iopp_2018}
E.~Ben-Sasson, I.~Bentov, Y.~Horesh, M.~Riabzev.
\newblock Fast Reed--Solomon Interactive Oracle Proofs of Proximity.
\newblock In \emph{ICALP 2018} (LIPIcs 107), pp.~14:1--14:17, 2018.
\newblock doi: \href{https://doi.org/10.4230/LIPIcs.ICALP.2018.14}{10.4230/LIPIcs.ICALP.2018.14}.

\hypertarget{r20}{}\bibitem{deep_fri_2019}
E.~Ben-Sasson, L.~Goldberg, S.~Kopparty, S.~Saraf.
\newblock DEEP-FRI: Sampling outside the box improves soundness.
\newblock arXiv:1903.12243, 2019.
\newblock \url{https://arxiv.org/abs/1903.12243}.

\hypertarget{r21}{}\bibitem{cryptoeprint:2018/046}
E.~Ben-Sasson, I.~Bentov, Y.~Horesh, M.~Riabzev.
\newblock Scalable, transparent, and post-quantum secure computational integrity.
\newblock Cryptology ePrint Archive, Paper 2018/046, 2018.
\newblock \url{https://eprint.iacr.org/2018/046}.

\hypertarget{r22}{}\bibitem{solana_commitment_levels}
Solana Labs.
\newblock Solana Commitment Status: Processed, Confirmed, Finalized, 2025.
\newblock accessed 2025-10-23.
\newblock \url{https://docs.solanalabs.com/consensus/commitments}.
\end{thebibliography}
\end{document}